# Expressions of Market-Based Correlations Between Prices and Returns of Two Assets

Victor Olkhov

Independent, Moscow, Russia

victor.olkhov@gmail.com

ORCID: 0000-0003-0944-5113

**Dec. 10, 2024**

**Abstract**

This paper derives the expressions of correlations between prices of two assets, returns of two assets, and price-return correlations of two assets that depend on statistical moments and correlations of the current values, past values, and volumes of their market trades. The usual frequency-based expressions of correlations of time series of prices and returns describe a partial case of our model when all trade volumes and past trade values are constant. Such an assumptions are rather far from market reality, and its use results in excess losses and wrong forecasts. Traders, banks, and funds that perform multi-million market transactions or manage billion-valued portfolios should consider the impact of large trade volumes on market prices and returns. The use of the market-based correlations of prices and returns of two assets is mandatory for them. The development of macroeconomic models and market forecasts like those being created by BlackRock's Aladdin, JP Morgan, and the U.S. Fed., is impossible without the use of market-based correlations of prices and returns of two assets.

This research received no support, specific grant or financial assistance from funding agencies in the public, commercial or nonprofit sectors. We welcome valuable offers of grants, support and positions.

# 1. Introduction

The correct assessments of correlations between prices and returns of two assets are important for portfolio selection, investors' expectations, forecasting of market trades, and the development of sophisticated macroeconomic models. This paper derives the expressions of correlations of prices and returns of two assets that reveal the dependence on the statistical moments and correlations of volumes, current, and past values of their market trades.

At least since Bachelier (1900), the description of prices as random variables was based on processing the time series of prices. The time series of prices and returns serve as the basis for the description of their statistical properties and for the assessments of their impact on economics, expectations, portfolio selection, etc. (Pearce, 1983; Cochrane and Hansen, 1992; Diebold and Yilmaz, 2009; Golez and Koudijs, 2017). Numerous studies describe correlations of the time series of asset prices, returns, and their forecasts: Eun and Resnick, (1984); Nelson and Kim, (1990); Knight and Satchell, (Ed). (2001); Andersen et al., (2006); Cochrane, (2006); Ferreira and Santa-Clara, (2008); Cortazar et al., (2016); Kelly et al., (2022). Any list of references can be significantly expanded. It is well known that market trade volumes impact prices and returns. The price-volume and return-volume relations were described by many researchers (Karpoff, 1987; Gallant, Rossi, and Tauchen, 1992; Campbell, Grossman, and Wang, 1993; Brock and LeBaron, 1995; Llorente et al., 2001; Dorn, Huberman, and Sengmueller, 2008; DeFusco, Nathanson and Zwick, 2022). In recent years, modeling and forecasting of time series of price and return began developed using LLM and AI (Gu, Kelly and Xiu, 2019; Cao, 2021; Kelly, and Xiu, 2023).

We highlight that all the above studies describe the random properties of prices and returns, their possible correlations, and their impact on financial markets and the economy based on processing the time series of prices and returns only. Simply speaking, it is generally assumed and accepted that time series of prices or returns deliver all necessary and sufficient information required for the assessments of their their averages, correlations, etc. Let us denote market prices $p(t_i|1)$ of asset 1 at time $t_i$ and $p(t_i\text{-}\beta|2)$ at time $t_i\text{-}\beta$ of asset 2. Let us assume that the time interval $\varepsilon$ between trades is constant and there were $N$ trades during the averaging interval $\Delta$ (1.1):

$$\Delta = \left[t - \frac{\Delta}{2}; t + \frac{\Delta}{2}\right] \; ; \; t_i \in \Delta \; ; \; t_i - t_{i-1} = \varepsilon \; ; \; i = 1,2,\ldots N \qquad (1.1)$$

For such a case, the usual expressions of frequency-based average price $\pi(t|1)$ of asset 1 and average price $\pi(t\text{-}\beta|2)$ of asset 2 take simple and familiar form (1.2):

$$\pi(t|1) = \frac{1}{N}\sum_{i=1}^{N} p(t_i|1) \; ; \; \pi(t - \beta|2) = \frac{1}{N}\sum_{i=1}^{N} p(t_i - \beta|2) \qquad (1.2)$$

Relations (1.2) approximate the mathematical expectations of prices by $N$ terms of time series during $\Delta$ (1.1). To estimate the frequency-based correlations $corr\{p(t|1),p(t-\beta|2)\}$ (1.3; 1.4) of prices $p(t_i|1)$ and $p(t_i-\beta|2)$ of two assets by $N$ terms of their time series, one should use familiar expressions (Shiryaev, 1999; Shreve, 2004):

$$corr\{p(t|1)p(t-\beta|2)\} = \frac{1}{N}\sum_{i=1}^{N}[p(t_i|1) - \pi(t|1)][p(t_i - \beta|2) - \pi(t-\beta|2)] \quad (1.3)$$

$$corr\{p(t|1), p(t-\beta|2)\} = \pi\pi(t,\beta|1,2) - \pi(t|1)\pi(t-\beta|2) \quad (1.4)$$

$$\pi\pi(t,\beta|1,2) = \frac{1}{N}\sum_{i=1}^{N} p(t_i|1)\, p(t_i - \beta|2) \quad (1.5)$$

The function $\pi\pi(t,\beta|1,2)$ denotes the frequency-based joint statistical moment of the product of prices of two assets (1.5). The expressions (1.2-1.5) are the basic for the evaluation of time series analysis of average prices and their correlations.

However, the time series of prices and returns are not the only sources of their random properties. At least since Berkowitz (1988) and subsequent studies (Duffie and Dworczak, 2018), volume-weighted-average-price (VWAP) gives the assessments of average price that are different from (1.2). In particular, VWAP is widely used for the definition of the average price (CME Group, 2024) at most exchanges as a tool to avoid fraud and manipulation of closing day price. To introduce VWAP, let us denote the value $C(t_i|1)$ of trade of asset 1 at time $t_i$ at price $p(t_i|1)$ with trade volume $U(t_i|1)$. The market trade value $C(t_i|1)$, volume $U(t_i|1)$, and price $p(t_i|1)$ follow a trivial trade price equation (1.6):

$$C(t_i|1) = p(t_i|1)U(t_i|1) \quad (1.6)$$

The frequency-based assessments of the *n-th* statistical moments of trade value $C(t;n|1)$ (1.7) and volume $U(t;n|1)$ (1.8) of asset 1 by $N$ terms of time series during $\Delta$ (1.1) take the form:

$$C(t|1) = \frac{1}{N}\sum_{i=1}^{N} C(t_i|1) \quad ; \quad C_\Sigma(t|1) = \sum_{i=1}^{N} C(t_i|1) = N \cdot C(t|1) \quad (1.7)$$

$$U(t|1) = \frac{1}{N}\sum_{i=1}^{N} U(t_i|1) \quad ; \quad U_\Sigma(t|1) = \sum_{i=1}^{N} U^n(t_i|1) = N \cdot U(t|1) \quad (1.8)$$

Functions $C_\Sigma(t|1)$ and $U_\Sigma(t|1)$ describe the total sums of trade values and volumes of asset 1 during $\Delta$ (1.1). The use of (1.6-1.8) allows define VWAP $a(t|1)$ (1.9):

$$a(t|1) = \frac{\sum_{i=1}^{N} p(t_i|1)U(t_i|1)}{\sum_{i=1}^{N} U(t_i|1)} = \frac{C_\Sigma(t|1)}{U_\Sigma(t|1)} = \frac{C(t|1)}{U(t|1)} \quad (1.9)$$

Actually, VWAP describes the average price $a(t|1)$ of asset 1 $p(t_i|1)$ weighted by the volumes $U(t_i|1)$ of market trades of asset 1 during the time interval $\Delta$ (1.1). The expression of VWAP $a(t|1)$ (1.9) as a ratio of total sums of trade values $C_\Sigma(t|1)$ (1.7) to sums of volumes $U_\Sigma(t|1)$ (1.8) during $\Delta$ (1.1) or equally the ratio of average value $C(t|1)$ (1.7) to average volume $U(t|1)$ (1.8) highlights its simple and clear economic meaning. One can easily notice that if all trade volumes $U(t_i|1)=U$, $i=1,..N$, are constant, the average price (1.9) takes its familiar frequency-based form (1.2). We use different notations for the frequency-based average price $\pi(t|1)$ (1.2) and for VWAP $a(t|1)$ (1.9) to highlight the distinctions between these definitions of the average price. VWAP $a(t|1)$ (1.9) reveals the dependence on the random properties of trade volumes $U(t_i|1)$, or equally, the dependence of the average price $a(t|1)$ (1.9) on the average value $C(t|1)$ (1.7) and volume $U(t|1)$ (1.8). Anyway, the introduction of VWAP $a(t|1)$ (1.9) highlights the duality of the description of random price average as VWAP $a(t|1)$ (1.9) or as frequency-based $\pi(t;1)$ (1.2) if one neglects the impact of random of trade volumes $U(t_i|1)$ and assumes that all trade volumes are constant.

Such a duality raises the question that correlations between prices of two assets could also be determined by different expressions. The conventional expression of the correlation (1.3-1.5) neglects the influence of the size and randomness of trade volumes. If one wants to consider the impact of random trade volumes $U(t_i|1)$ of asset 1 and of $U(t_i-\beta|2)$ of asset 2, then the expression of correlations should take a different market-based form. For that case, the use of price time series alone is not sufficient for the definition of correlations. The market origin of price (1.6) requires to consider correlations of the values and volumes of market trades as the cause that determines correlations of market prices and returns.

The consideration of the influence of random trading values and volumes on the average prices and correlations between prices of two assets is mandatory for those traders, investors, and institutions that make multi-million market transactions that impact the market price dynamics. The selections of portfolios that are collected of big volumes of assets with great market values are impossible without the assessments of correlations of prices and returns that take into account the influence of large market trades. The macroeconomic models and forecasts of market trades unworkable without the use of the dependence of averages, volatilities, and correlations of market prices and returns on random values and volumes of market trades. We believe that BlackRock's Aladdin, JP Morgan, and U.S. Fed macroeconomic economic modeling and market forecasting should pay the attention to the influence of trade volumes on the volatilities and correlations of prices and returns.

Our paper derives market-based expressions of correlations between prices of two assets, returns of two assets, and price-return correlations of two assets that discover their dependence on the statistical moments and correlations of trades of two assets.

In Section 2 and App. A, we consider market-based expressions of correlations of prices of two assets. In Section 3 and App. B, we derive market-based correlations of returns of two assets. In Section 4 and App. C., we present market-based correlations of prices of asset 1 and returns of asset 2. Conclusion is in Section 5. In this paper, we assume that all prices are adjusted to the current time *t*. This paper follows the description of market-based statistical moments of prices and returns, derived by Olkhov (2022-2024), and we refer there for details. We believe that readers know or can find on their own the meaning of notations and methods that are not given in the text.

## 2. Market-based correlations of prices of two assets

We give the detailed derivation of a market-based expression of the correlation of prices of two assets in App. A. Here, we present a brief consideration.

The trade price equation (1.6) generates VWAP (1.9). To derive the expression for correlations between two assets, we consider the product (2.2) of equation (1.6) that describes trade values, volumes, and prices of asset 1 and equation (2.1; A.8) that describes asset 2:

$$C(t_i - \beta|2) = p(t_i - \beta|2)U(t_i - \beta|2) \qquad (2.1)$$

$$C(t_i|1)C(t_i - \beta|2) = p(t_i|1)\, p(t_i - \beta|2)\, U(t_i|1)U(t_i - \beta|2) \qquad (2.2)$$

The form of equation (2.2) is the same as (1.6), and we show how it can generate the market-based expression of correlations between market prices *p(t_i|1)* of asset *1* and prices *p(t_i-β|2)* of asset *2*. Relations (2.3) present the familiar definition of correlation *corr{p(t|1),p(t-β|2)}* via market-based mathematical expectation *E_m[..]*:

$$corr\{p(t|1), p(t - \beta|2)\} = E_m\big[[p(t_i|1) - a(t|1)][p(t_i - \beta|2) - a(t - \beta|2)]\big] \qquad (2.3)$$

As *E_m[..]* we denote market-based averaging over weight functions (A.14; A.15). We highlight that contrary to frequency-based correlation (1.3), market-based correlation (2.3) of prices of two assets is calculated with respect to VWAP *a(t|1)* (A.9) of asset 1 and VWAP *a(t-β|2)* (A.10) of asset 2 that differ from frequency-based average prices *π(t|1)* of asset 1 and *π(t-β|2)* of asset 2 (1.2). Finally, as we show in App. A., the correlation *corr{p(t|1),p(t-β|2)}* (2.3) between prices of two assets takes the form (2.4; A.32):

$$corr\{p(t|1), p(t-\beta|2)\} = \frac{1}{UU(t,\beta|1,2)}[corr\{C(t|1), C(t-\beta|2)\} - a(t|1)corr\{U(t|1), C(t-\beta|2)\} - a(t-\beta|2)corr\{C(t|1), U(t-\beta|2)\} + a(t|1)a(t-\beta|2)corr\{U(t|1), U(t-\beta|2)\}] \qquad (2.4)$$

The definitions of functions and notations are given in (A.9; A.10), (A.19-A.26). If all trade volumes *U(t_i|1)* and *U(t_i-β|2)* of both assets are constant during the averaging interval *Δ* (1.1), the expression of correlation takes the conventional form (1.2-1.5) of frequency-based correlation of prices of two assets. The obvious distinctions between the frequency-based expression of correlations (1.2-1.5) of two time series of prices of two assets and the expression of market-based correlations of prices of two assets (A.32) discover the impact of correlations of market trade values and volumes. We underline that correlations between

values and volumes of market trades of two assets are determined by frequency-based expressions (A19-A.26).

If assets 1 and 2 are the same, then the correlation of two assets (A.32) takes the form of a market-based correlation (2.5; A.33) of prices of a single asset 1.

$$corr\{p(t|1)p(t-\beta|1)\} = \frac{1}{UU(t,\beta|1)}[corr\{C(t|1),C(t-\beta|1)\} - a(t|1)corr\{U(t|1),C(t-\beta|1)\} - a(t-\beta|1)corr\{C(t|1),U(t-\beta|1)\} + a(t|1)a(t-\beta|1)corr\{U(t|1),U(t-\beta|1)\}] \quad (2.5)$$

Finally, if the time shift $\beta$ between prices $\beta=0$, then the correlation of two prices (A.33) equals to market-based volatility $\sigma_p^2(t)$ (2.6; A.35) of price at time $t$:

$$\sigma_p^2(t) = \frac{1}{U(t;2|1)}[\Omega_C^2(t) - 2a(t|1)corr\{C(t|1),U(t|1)\} + a^2(t|1)\Omega_U^2(t)] \quad (2.6)$$

Functions $\Omega_C^2(t)$ and $\Omega_U^2(t)$ (A.36) determine volatilities of trade values $C(t_i|1)$ and trade volumes $U(t_i|1)$ during the averaging interval $\Delta$ (1.1).

## 3. Market-based correlations of returns of two assets

We present the detailed derivation in App. B. We denote return $r(t_i,\alpha|1)$ (3.1) of asset 1 at time $t_i$ with the shift $\alpha$ and return $r(t_i,\beta|2)$ (3.1) of asset 2 at time $t_i$ with the shift $\beta$ as:

$$r(t_i,\alpha|1) = \frac{p(t_i|1)}{p(t_i-\alpha|1)} \quad ; \quad r(t_i,\beta|2) = \frac{p(t_i|2)}{p(t_i-\beta|2)} \quad (3.1)$$

Like relations (1.2-1.5) for price time series, the time series of returns $r(t_i,\alpha|1)$ and $r(t_i,\beta|2)$ (3.1) are the basic source for the definitions of frequency-based average returns (3.2):

$$\varrho(t,\alpha|1) = \frac{1}{N}\sum_{i=1}^{N}r(t_i,\alpha|1) \quad ; \quad \varrho(t,\beta|2) = \frac{1}{N}\sum_{i=1}^{N}r(t_i,\beta|2) \quad (3.2)$$

The expressions of frequency-based correlation $corr\{r(t,\alpha|1)),r(t,\beta|2)\}$ (3.3-3.5) of returns $r(t_i,\alpha|1)$ and $r(t_i,\beta|2)$ (3.1) by $N$ terms of market trades have the form similar to (1.3):

$$corr\{r(t,\alpha|1),r(t,\beta|2)\} = \frac{1}{N}\sum_{i=1}^{N}[r(t_i,\alpha|1) - \varrho(t,\alpha|1)][r(t_i,\beta|2) - \varrho(t,\beta|2)] \quad (3.3)$$

$$corr\{r(t,\alpha|1),r(t,\beta|2)\} = \varrho\varrho(t,\alpha,\beta|1,2) - \varrho(t,\alpha|1)\varrho(t,\beta|2) \quad (3.4)$$

$$\varrho\varrho(t,\alpha,\beta|1,2) = \frac{1}{N}\sum_{i=1}^{N}r(t_i,\alpha|1)\,r(t_i,\beta|2) \quad (3.5)$$

The function $\rho\rho(t,\alpha,\beta|1,2)$ denotes the frequency-based joint statistical moment of the product of returns $r(t_i,\alpha|1)$ and $r(t_i,\beta|2)$ (3.1). The doubleness of definitions of frequency-based averages and market-based averages of return and corresponding doubleness of expressions of frequency-based correlations and market-based correlations of returns of two assets is completely similar to the doubleness of definitions of correlations of prices.

However, volume weighted average returns don't exist. What one can consider as a definition of market-based average return that is alike to VWAP? Actually, 36 years before Berkowitz et al. (1988) introduced VWAP, in 1952, H. Markowitz published his famous paper titled "Portfolio Selection." Markowitz considered the portfolio composed of $N$ securities with returns $r_i$, $i=1,..N$, and defined the return of the portfolio as "weighted with weights equal the relative amount invested in security" (Markowitz, 1952). Actually, that definition has the form that completely coincides with the form of VWAP. Let denote $X_i$ as investment into security $i$, $i=1,2,..N$, and present the return $R$ of the portfolio as (3.6):

$$R = \frac{1}{\sum_{i=1}^{N} X_i} \sum_{i=1}^{N} r_i X_i = \sum_{i=1}^{N} r_i x_i \quad ; \quad x_i = \frac{X_i}{\sum_{i=1}^{N} X_i} \tag{3.6}$$

In (3.6), $x_i$ – is a "relative amount invested in security." We remind you that all prices are adjusted to current time $t$. Actually, there is no difference between the consideration of the return of the portfolio that is composed of $N$ securities and the consideration of the average return of $N$ market trades during $\Delta$ (1.1). To show that, let us use (3.1) and transform the trade price equations (1.6) and (2.1) into trade return equations (3.7; B.2; 3.8; B.3):

$$C(t_i|1) = p(t_i|1)U(t_i|1) = \frac{p(t_i|1)}{p(t_i - \alpha|1)} p(t_i - \alpha|1)U(t_i|1)$$

$$C(t_i|1) = r(t_i,\alpha|1)C_0(t_i,\alpha|1) \quad ; \quad C_0(t_i,\alpha|1) = p(t_i - \alpha|1)U(t_i|1) \tag{3.7}$$

$$C(t_i|2) = r(t_i,\beta|2)C_0(t_i,\beta|2) \quad ; \quad C_o(t_i,\beta|2) = p(t_i - \beta|2)U(t_i|2) \tag{3.8}$$

The function $C_0(t_i,\alpha|1)$ (3.7) equals the value of the current trade volume $U(t_i|1)$ of asset 1 in the past at time $t_i$-$\alpha$ at price $p(t_i$-$\alpha|1)$. Respectively, $C_0(t_i,\beta|2)$ (3.8) equals to the value of trade volume $U(t_i|2)$ of the asset 2 in the past at time $t_i$-$\beta$ at price $p(t_i$-$\beta|2)$. One can consider $N$ trades during $\Delta$ (1.1) as the portfolio of $N$ securities, and consider the past values $C_0(t_i,\alpha|1)$ (3.7) and $C_0(t_i,\beta|2)$ (3.8) as *the amounts invested in security i*. One can consider the definition of return $R$ (3.6) of the portfolio of $N$ securities as the definition of Value-Weighed-Average-Return (VaWAR) $h(t,\alpha|1)$ (3.9; B.6) and $h(t,\beta|2)$ (3.10; B.7) of $N$ trades of assets 1 and 2:

$$h(t,\alpha|1) = \frac{\sum_{i=1}^{N} r(t_i,\alpha|1)C_o(t_i,\alpha|1)}{\sum_{i=1}^{N} C_o(t_i,\alpha|1)} = \frac{C_\Sigma(t|1)}{C_{o\Sigma}(t,\alpha|1)} = \frac{C(t|1)}{C_o(t,\alpha|1)} \tag{3.9}$$

$$h(t,\beta|2) = \frac{\sum_{i=1}^{N} r(t_i,\beta|2)C_o(t_i,\beta|2)}{\sum_{i=1}^{N} C_o(t_i,\beta|2)} = \frac{C_\Sigma(t|2)}{C_{o\Sigma}(t,\beta|2)} = \frac{C(t|2)}{C_o(t,\beta|2)} \tag{3.10}$$

Average past values $C_0(t,\alpha|1)$ (3.11; B.4) of asset 1 and $C_0(t,\beta|2)$ (3.12; B.5) of asset 2 take the frequency-based form:

$$C_o(t,\alpha|1) = \frac{1}{N}\sum_{i=1}^{N} C_o(t_i,\alpha|1) \quad ; \quad C_{o\Sigma}(t,\alpha|1) = N \cdot C_o(t,\alpha|1) \tag{3.11}$$

$$C_o(t,\beta|2) = \frac{1}{N}\sum_{i=1}^{N} C_o(t_i,\beta|2) \quad ; \quad C_{o\Sigma}(t,\beta|2) = N \cdot C_o(t,\beta|2) \tag{3.12}$$

The functions $C_{0\Sigma}(t,\alpha|1)$ (3.11) and $C_{0\Sigma}(t,\beta|2)$ (3.12) denote the total sums of past values during $\Delta$ (1.1). "*The relative amount invested in security*" $x_i$ (3.6) takes form of relative past values $c_0(t_i,\alpha|1)$ of asset 1 and $c_0(t_i,\beta|2)$ of asset 2 of trades (3.13):

$$c_o(t_i,\alpha|1) = \frac{C_o(t_i,\alpha|1)}{\sum_{i=1}^{N} C_o(t_i,\alpha|1)} \quad ; \quad c_o(t_i,\beta|2) = \frac{C_o(t_i,\beta|2)}{\sum_{i=1}^{N} C_o(t_i,\beta|2)} \quad (3.13)$$

The dualism between the definitions of frequency-based average returns $\rho(t,\alpha|1)$ and $\rho(t,\beta|2)$ (3.2) and VaWAR $h(t,\alpha|1)$ (3.9) and $h(t,\beta|2)$ (3.10) is completely the same as the dualism between the definitions of frequency-based average prices $\pi(t|1)$ and $\pi(t-\beta|2)$ (1.2) and VWAP $a(t|1)$ (1.9; A.9) and $a(t-\beta|2)$ (A.10). In the case that all past values $C_0(t_i,\alpha|1)$ (3.7) and $C_0(t_i,\beta|2)$ (3.8) of *N* trades are constant during $\Delta$ (1.1), then VaWAR $h(t,\alpha|1)$ (3.9) and $h(t,\beta|2)$ (3.10) become equal to frequency-based average returns $\rho(t,\alpha|1)$ and $\rho(t,\beta|2)$ (3.2).

We highlight that the economic assumptions determine the choice of probability of returns. If investors assume that the "relative amount invested in security" is proportional to *1/N*, and *N* is the number of securities, or equally, if investors assume that the past values of all trades are constant, then the frequency-based assessments (3.2) determine the average return during $\Delta$ (1.1). However, if investors consider the influence of the "relative amount invested in security", then the return of the portfolio takes the form determined by Markowitz and the market-based average return of *N* terms of trade time series takes the form of VaWAR (3.9; 3.10). The choice between these two assumptions determines the distinctions between the expressions of correlations of returns of two assets. The expressions (3.3-3.5) describe correlations that neglect the randomness of the "relative amounts invested in security".

To derive the market-based expression of correlation of returns of two assets that depends on statistical moments and correlations of current and past trade values, we follow the logic of the derivation of correlations of prices of two assets and consider the product (3.14; B.8) of return equations (3.7) and (3.8):

$$C(t_i|1)C(t_i|2) = r(t_i,\alpha|1)r(t_i,\beta|2)C_0(t_i,\alpha|1)C_o(t_i,\beta|2) \quad (3.14)$$

The market-based correlation (3.15; B.11) of returns of two assets has the form like (2.3):

$$corr\{r(t,\alpha|1), r(t,\beta|2)\} = E_m[[r(t_i,\alpha|1) - h(t,\alpha|1)][r(t_i,\beta|2) - h(t,\beta|2)]] \quad (3.15)$$

We underline that contrary to the frequency-based correlation (3.3) that is calculated with respect to the frequency-based average returns $\rho(t,\alpha|1)$ and $\rho(t,\beta|2)$ (3.2), correlation (3.15) is calculated with respect to VaWAR $h(t,\alpha|1)$ (3.9) and $h(t,\beta|2)$ (3.10). Market-based mathematical expectation $E_m[..]$ in (3.15) is determined as averaging by the weight function

$$z(t_i,\alpha,\beta|1,2) = \frac{C_0(t_i,\alpha|1)C_o(t_i,\beta|2)}{\sum_{i=1}^{N} C_0(t_i,\alpha|1)C_o(t_i,\beta|2)} \quad ; \quad \sum_{i=1}^{N} z(t_i,\alpha,\beta|1,2) = 1 \quad (3.16)$$

The economic meaning of the weight function (3.16) is the relative amount of the product of past values $C_0(t_i,\alpha|1)$ (3.7) and $C_0(t_i,\beta|2)$ (3.8) of assets 1 and 2. It is alike to the relative past value (3.13) and has a similar meaning as Markowitz's weight function that equals to the

"relative amount invested in security." We give the detailed derivation of the dependence of the correlation of returns of two assets on statistical moments and correlations of their current and past values in App. B. The final form of the correlation of returns (B.24; 3.17):

$$corr\{r(t,\alpha|1), r(t,\beta|2)\} = \frac{1}{C_o C_o(t,\alpha,\beta|1,2)} [corr\{C(t|1), C(t|2)\} -$$

$$h(t,\beta|2) corr\{C(t|1), C_o(t,\beta|2)\} - h(t,\alpha|1) corr\{C_o(t,\alpha|1), C(t|2)\} +$$

$$h(t,\alpha|1) h(t,\beta|2) \, corr\{C_o(t,\alpha|1), C_o(t,\beta|2)\}] \quad (3.17)$$

The definitions of all notations are given in App. B. If one assumes that all past values $C_0(t_i,\alpha|1)$ (3.7) and $C_0(t_i,\beta|2)$ (3.8) of two assets are constant during $\Delta$ (1.1), then (see App. B.) correlation (3.17) between returns of two assets takes the form of frequency-based correlation (3.3; 3.4) of two time series of returns of two assets.

If assets 1 and 2 are the same, then the relations (3.17) describe autocorrelation $corr\{r(t,\alpha|1), r(t,\beta|1)\}$ (3.18; B.25) between returns of asset 1 with time shifts $\alpha$ and $\beta$.

$$corr\{r(t,\alpha|1), r(t,\beta|1)\} = \frac{1}{C_o C_o(t,\alpha,\beta|1)} [\Omega_C^2(t) - h(t,\beta|1) corr\{C(t|1), C_o(t,\beta|1)\} -$$

$$h(t,\alpha|1) corr\{C_o(t,\alpha|1), C(t|1)\} + h(t,\alpha|1) h(t,\beta|1) corr\{C_o(t,\alpha|1), C_o(t,\beta|1)\}] \quad (3.18)$$

If $\alpha=\beta$, (3.18) equals to the volatility of return $\sigma_r^2(t,\alpha)$ (3.19; B.26) with time shift $\alpha$:

$$\sigma_r^2(t,\alpha) = \frac{\Omega_C^2(t) + h^2(t,\alpha|1)\Phi^2(t,\alpha) - 2h(t,\alpha|1) corr\{C(t|1), C_o(t,\alpha|1)\}}{C_o(t,\alpha;2|1)} \quad (3.19)$$

The function $\Omega_C^2(t)$ denotes volatility (A.38) of trade values $C(t_i|1)$, and $\Phi^2(t,\alpha)$ denotes volatility (3.20; B.27) of past trade values:

$$\Phi^2(t,\alpha) = C_0(t,\alpha;2|1) - C_0^2(t,\alpha|1) \quad (3.20)$$

$$C_0(t,\alpha;2|1) = \frac{1}{N}\sum_{i=1}^{N} C_0^2(t_i,\alpha|1)$$

The function $C_0(t,\alpha;2|1)$ in (3.19; 3.20) has the meaning of the 2nd statistical moment or the average squares of past values $C_0(t_i,\alpha|1)$ of asset 1. The expression of market-based volatility of return $\sigma_r^2(t,\alpha)$ (3.19) coincides with the one derived by Olkhov (2023a; 2024).

The market-based volatility $\sigma_r^2(t,\alpha)$ (3.19) of returns of asset 1 reveals the complex dependence on volatilities on VaWAR returns $h(t,\alpha|1)$, current and past trade values $\Omega_C^2(t)$ and $\Phi^2(t,\alpha)$, and their correlation $corr\{C(t|1), C_0(t,\alpha|1)\}$, and on the 2nd statistical moment of past values $C_0(t_i,\alpha|1)$ of asset 1.

## 4. Market-based correlations of prices of asset 1 and returns of asset 2

To derive the dependence of the market-based correlation of prices and returns of two assets, one should use the results of sections 2 and 3. We give the detailed derivation in App.C. Similar to frequency-based expressions of correlation of prices of two assets (1.3-1.5) and the expressions of correlation of returns of two assets (3.3-3.5), the frequency-based expressions of correlation of prices and returns of assets 1 and 2 take the form:

$$corr\{p(t|1), r(t,\beta|2)\} = \frac{1}{N}\sum_{i=1}^{N}[p(t_i|1) - \pi(t|1)][r(t_i,\beta|2) - \varrho(t,\beta|2)] \quad (4.1)$$

$$corr\{p(t|1), r(t,\beta|2)\} = \pi\varrho(t,\beta|1,2) - \pi(t|1)\varrho(t,\beta|2) \quad (4.2)$$

$$\pi\varrho(t,\beta|1,2) = \frac{1}{N}\sum_{i=1}^{N} p(t_i|1)\, r(t_i,\beta|2) \quad (4.3)$$

To derive a market-based expression of the correlation of prices of asset 1 and returns of asset 2 that reflect the dependence on statistical moments and correlations of current trade values of assets, trade volumes of asset 1, and past values of asset 2, we define the product (4.4; C.1) of the trade price equation (1.6) and return equation (3.8):

$$C(t_i|1)C(t_i|2) = p(t_i|1)r(t_i,\beta|2)U(t_i|1)C_o(t_i,\beta|2) \quad (4.4)$$

We define market-based correlation of prices and returns of two assets as (4.5):

$$corr\{p(t|1), r(t,\beta|2)\} = E_m\big[[p(t_i|1) - a(t|1)][r(t_i,\beta|2) - h(t,\beta|2)]\big] \quad (4.5)$$

Market-based mathematical expectation $E_m[..]$ in this case is determined by the joint weigh function $\psi(t_i,\beta|1,2)$ (C.2; 4.6):

$$\psi(t_i,\beta|1,2) = \frac{U(t_i|1)C_o(t_i,\beta|2)}{\sum_{i=1}^{N} U(t_i|1)C_o(t_i,\beta|2)} \quad (4.6)$$

The joint weigh function $\psi(t_i,\beta|1,2)$ (4.6) has meaning of the relative weight of product of trade volume $U(t_i|1)$ of asset 1 and past value $C_0(t_i,\beta|2)$ of asset 2. The calculations (App.C) give the final dependence (4.7; C.9) of market-based correlation of prices and returns:

$$corr\{p(t|1), r(t,\beta|2)\} = \frac{1}{UC_o(t,\beta|1,2)}\big[corr\{C(t|1), C(t|2)\} -$$
$$h(t,\beta|2)corr\{C(t|1), C_o(t,\beta|2)\} - a(t|1)corr\{U(t|1), C(t|2)\} +$$
$$a(t|1)h(t,\beta|2)corr\{U(t|1), C_o(t,\beta|2)\}\big] \quad (4.7)$$

One can check that if all trade volumes $U(t_i|1)$ of asset 1 and past values $C_0(t_i,\beta|2)$ of asset 2 are constant, then relations (4.7) take the form (4.1-4.3) of a frequency-based expression of correlation between time series of prices of asset 1 and returns of asset 2.

## 5. Conclusion

We derived the expressions of market-based of correlations between prices and returns of two assets that reveal their complex dependence on statistical moments and correlations of current and past values and volumes of their market trades. The correlations are determined with respect to market-based average price as VWAP, and average return as VaWAR.

If one assumes that all trade volumes and past values of both assets are constant during the averaging interval, all expressions of correlations take the form of usual frequency-based correlations between time series of prices and returns of two assets. Those market participants who perform multi-million transactions and manage billion-valued portfolios, should keep in mind that the frequency-based assessments of correlations use approximations of constant volumes and constant past values of market trades. These non-market approximations could lead to huge losses. The neglecting of the randomness of trade volumes and past values in macroeconomic modeling and market forecasting could result in significantly wrong predictions.

The complexity of market-based expressions of correlations matches the complexity of the corresponding economic problems and the description of market trade. Modeling and predictions of market-based correlations require forecasting of the trade volumes, current and past trade values, their averages and correlations. That significantly complicates the predictions of averages and correlations of prices and returns. Such forecasts require huge market data, high expenses, and sophisticated economic and IT models. Primarily, our results would benefit the largest macroeconomic modelers like BlackRock's Aladdin, JP Morgan, or the U.S. Fed. Investors and institutions that manage multi-billion portfolios should consider our results to avoid unexpected losses as a result of miscalculations of correlations of their assets. The use of market-based correlations and their various approximations and simplifications would be useful for academic researchers that study real market dynamics and model market portfolios.

# Appendix A
## Market-Based Correlations of Prices of Two Assets

This Appendix describes the dependence of market-based correlations of prices of two assets during the averaging interval $\Delta$ on statistical moments and correlations of values and volumes of their market trades. We assume that $N$ market trades with assets $1$ and $2$ were made with a constant interval $\varepsilon$ between trades at time $t_i$:

$$\varepsilon < \Delta \quad ; \quad \Delta = \left[t - \frac{\Delta}{2}; t + \frac{\Delta}{2}\right] \quad ; \quad t_i = t_{i-1} + \varepsilon \in \Delta \quad ; \quad i = 1,2,\ldots N \quad (A.1)$$

Let us denote the trade values, volumes, and prices of trades at time $t_i$ with asset $1$ and at time $t_i$-$\beta$ respectively:

$$C(t_i|1), \ U(t_i|1), \ p(t_i|1) \quad ; \quad C(t_i - \beta|2), \ U(t_i - \beta|2), \ p(t_i - \beta|2) \quad (A.2)$$

We assume that $t_i$-$\beta$ belongs to the time series of trades in the past. $N$ terms of time series of trade values $C(t_i|1)$, $C(t_i$-$\beta|2)$ and volumes $U(t_i|1)$, $U(t_i$-$\beta|2)$ approximate their averages:

$$C(t|1) = \frac{1}{N}\sum_{i=1}^{N} C(t_i|1) \quad ; \quad U(t|1) = \frac{1}{N}\sum_{i=1}^{N} U(t_i|1) \tag{A.3}$$

$$C(t-\beta|2) = \frac{1}{N}\sum_{i=1}^{N} C(t_i-\beta|2) \quad ; \quad U(t-\beta|2) = \frac{1}{N}\sum_{i=1}^{N} U(t_i-\beta|2) \tag{A.4}$$

We repeat (1.7; 1.8) for convenience. We denote $C_\Sigma(t|1)$, $C_\Sigma(t-\beta|2)$, $U_\Sigma(t|1)$, $U_\Sigma(t-\beta|2)$ – the total sums of the values and volumes of assets *1* and *2*:

$$C_\Sigma(t|1) = N \cdot C(t|1) = \sum_{i=1}^{N} C(t_i|1) \quad ; \quad C_\Sigma(t-\beta|2) = N \cdot C(t-\beta|2) \tag{A.5}$$

$$U_\Sigma(t|1) = N \cdot U(t|1) = \sum_{i=1}^{N} U(t_i|1) \quad ; \quad U_\Sigma(t-\beta|2) = N \cdot U(t-\beta|2) \tag{A.6}$$

The values *C(t$_i$|1)* and volumes *U(t$_i$|1)* of market trades at *t$_i$* define the market price *p(t$_i$|1)* of asset 1 due to a trivial equation (A.7):

$$C(t_i|1) = p(t_i|1)U(t_i|1) \tag{A.7}$$

Similar equation (A.8) defines the price *p(t$_i$-β|2)* of asset 2 by the market trade at time *t$_i$-β*

$$C(t_i - \beta|2) = p(t_i - \beta|2)U(t_i - \beta|2) \tag{A.8}$$

The equations (A.7; A.8) generate the familiar definitions of VWAP (Berkowitz et al., 1988; Duffie Dworczak, 2018). For assets 1 and 2, VWAP *a(t|1)* and *a(t- β|2)* take the form:

$$E_m[p(t_i|1)] = a(t|1) = \frac{\sum_{i=1}^{N} p(t_i|1)U(t_i|1)}{\sum_{i=1}^{N} U(t_i|1)} = \frac{C_\Sigma(t|1)}{U_\Sigma(t|1)} = \frac{C(t|1)}{U(t|1)} \tag{A.9}$$

$$E_m[p(t_i - \beta|2)] = a(t - \beta|2) = \frac{\sum_{i=1}^{N} p(t_i-\beta|2)U(t_i-\beta|2)}{\sum_{i=1}^{N} U(t_i-\beta|2)} = \frac{C_\Sigma(t-\beta|2)}{U_\Sigma(t-\beta|2)} = \frac{C(t-\beta|2)}{U(t-\beta|2)} \tag{A.10}$$

We denote VWAP (A.9; A.10) as market-based averages $E_m[p(t_i|1)]=a(t|1)$ (A.9) and $E_m[p(t_i-\beta|2)]=a(t-\beta|2)$ (A.10) of prices *p(t$_i$|1)* and *p(t$_i$-β|2)* of assets 1 and 2 during *Δ*. The VWAP (A.9) has meaning of the ratio of average value *C(t|1)* to average volume *U(t|1)* (A.3) or the ratio of total value *C$_\Sigma$(t|1)* (A.5) to total volume *U$_\Sigma$(t|1)* (A.6). VWAP *a(t-β|2)* (A.10) of the asset 2 at time *t-β* has the same meaning. VWAP or market-based averages *a(t|1)* (A.9) of the asset 1 and *a(t-β|2)* (A.10) of the asset 2 differ from the mean prices *π(t|1)* and *π(t-β|2)* (A.11) of assets 1 and 2 that are determined by frequency-based averaging of price time series *p(t$_i$|1)* (A.7) and *p(t$_i$-β|2)* (A.8) during *Δ* (1.1; A.1):

$$\pi(t|1) = \frac{1}{N}\sum_{i=1}^{N} p(t_i|1) \quad ; \quad \pi(t-\beta|2) = \frac{1}{N}\sum_{i=1}^{N} p(t_i - \beta|2) \tag{A.11}$$

These distinctions highlight the dependence of VWAP or market-based averages (A.9; A.10) on random trade values *U(t$_i$|1)* and *U(t$_i$-β|2)* with assets 1 and 2. To derive market-based correlation *corr{p(t|1)p(t-β|2)}* between prices of assets 1 and 2 that takes into account VWAP (A.9; A.10) and reveal the dependence on statistical moments and correlations of trade values and volumes, let us consider the product of equations (A.7; A.8):

$$C(t_i|1)C(t_i - \beta|2) = p(t_i|1) p(t_i - \beta|2) U(t_i|1)U(t_i - \beta|2) \tag{A.12}$$

Equation (A.12) is similar to (A.7) or (A.8). Equations (A.7; A.8) determine the weight functions $u(t_i|1)$ and $u(t_i-\beta|2)$ (A.13) that define VWAP $a(t|1)$ (A.9) and $a(t-\beta|2)$ (A.10).

$$u(t_i|1) = \frac{U(t_i|1)}{U_\Sigma(t|1)} \; ; \; u(t_i - \beta|2) = \frac{U(t_i-\beta|2)}{U_\Sigma(t-\beta|2)} \; ; \; \sum_{i=1}^{N} u(t_i|1) = \sum_{i=1}^{N} u(t_i - \beta|2) = 1 \quad (A.13)$$

Equation (A.12) defines the weight function $w(t_i,\beta|1,2)$ (A.14) that has the form similar to weight functions $u(t_i|1)$ and $u(t_i-\beta|2)$ (A.13):

$$w(t_i, \beta|1,2) = \frac{U(t_i|1)U(t_i-\beta|2)}{UU_\Sigma(t,\beta|1,2)} \; ; \; \sum_{i=1}^{N} w(t_i, \beta|1,2) = 1 \quad (A.14)$$

$$UU_\Sigma(t,\beta|1,2) = \sum_{i=1}^{N} U(t_i;1)U(t_i - \beta|2) \quad (A.15)$$

To derive the expression of market-based correlation $corr\{p(t|1)p(t-\beta|2)\}$ we introduce the product $\delta p(t_i,\beta|1,2)$ (A.16) of variations of prices $p(t_i|1)$ and $p(t_i-\beta|2)$ of assets 1 and 2 with respect to their market-based average values $a(t|1)$ (A.9) and $a(t-\beta|2)$ (A.10).

$$\delta p(t_i, \beta|1,2) = [p(t_i|1) - a(t|1)][p(t_i - \beta|2) - a(t - \beta|2)] \quad (A.16)$$

We define $corr\{p(t|1),p(t-\beta|2)\}$ as market-based mathematical expectation of the product $\delta p(t_i,\beta|1,2)$ (A.16) as the averaging of $\delta p(t_i,\beta|1.2)$ by the weight functions $w(t_i,\beta|1,2)$ (A.14):

$$corr\{p(t|1), p(t - \beta|2)\} = E_m[\delta p(t_i, \beta|1,2)] = \sum_{i=1}^{N} \delta p(t_i, \beta; 1,2) w(t_i, \beta|1,2) \quad (A.17)$$

We highlight that our definitions of the market-based mathematical expectations $E_m[..]$ reveal averaging of prices $p(t_i|1)$ and $p(t_i-\beta|2)$, and averaging of correlations $E_m[\delta p(t_i,\beta|1,2)]=corr\{p(t|1),p(t-\beta|2)\}$ (A.17) by different weigh functions $u(t_i|1)$ and $u(t_i-\beta|2)$ (A.13) and by $w(t_i,\beta|1,2)$ (A.14) respectively. There are great differences between the notions of weight functions and probability measures. The probability of a random variable can be determined by the sequence of the *n-th* statistical moments (Shephard, 1991; Shiryaev, 1999; Shreve, 2004), and different statistical moments can be determined by the averaging over different weight functions. We approximate probability of prices by a finite set of statistical moments in Olkhov (2022a; 2023a). As we show below, the definition (A.17) provides the consistency of the 2$^{nd}$ joint statistical moment $E_m[p(t_i|1)p(t_i-\beta|2]$ that depends on averaging over the weight function $w(t_i,\beta|1,2)$ (A.14) with the 1$^{st}$ statistical moments $a(t|1)$ and $a(t-\beta|2)$ (A.9; A.10) that are averaged over the weight functions $u(t_i|1)$ and $u(t_i-\beta|2)$ (A.13). In addition, (A.17) guarantees that for the case $\beta=0$ and the assets *1* and *2* are the same the market-based volatility of price $\sigma_p^2(t)$ (A.18) is always non-negative. We refer Olkhov (2022a; 2022b) for details.

$$corr\{p(t|1), p(t - \beta|2)\}|_{\beta=0;1=2} = E_m[(p(t|1) - a(t|1))^2] = \sigma_p^2(t) \geq 0 \quad (A.18)$$

To calculate $E_m[\delta p(t_i,\beta|1,2)]=corr\{p(t|1),p(t-\beta|2)\}$ (A.17) we introduce the joint statistical moments of trade values and volumes of assets 1 and 2:

$$CC(t,\beta|1,2) = \frac{1}{N}\sum_{i=1}^{N} C(t_i; 1)C(t_i - \beta|2) \quad ; \quad CC_{\Sigma}(t,\beta|1,2) = N \cdot CC(t,\beta|1,2) \quad (A.19)$$

$$UU(t,\beta|1,2) = \frac{1}{N}\sum_{i=1}^{N} U(t_i; 1)U(t_i - \beta|2) \quad ; \quad UU_{\Sigma}(t,\beta|1,2) = N \cdot UU(t,\beta|1,2) \quad (A.20)$$

$$CU(t,\beta|1,2) = \frac{1}{N}\sum_{i=1}^{N} C(t_i; 1)U(t_i - \beta|2) \quad ; \quad CU_{\Sigma}(t,\beta|1,2) = N \cdot CU(t,\beta|1,2) \quad (A.21)$$

$$UC(t,\beta|1,2) = \frac{1}{N}\sum_{i=1}^{N} U(t_i; 1)C(t_i - \beta|2) \quad ; \quad UC_{\Sigma}(t,\beta|1,2) = N \cdot UC(t,\beta|1,2) \quad (A.22)$$

Function $CC(t,\beta|1,2)$ (A.19) describes the approximation of the joint average or joint statistical moment of the trade values $C(t_i|1)$ of asset 1 and trade values $C(t_i-\beta|2)$ of asset 2 by $N$ terms of their trades during the averaging interval $\Delta$ (A.1). Function $CC_{\Sigma}(t,\beta|1,2)$ (A.19) describes the total sum of the product $C(t_i|1)C(t_i-\beta|2)$ during the interval $\Delta$ (A.1). Other functions (A.20-A.22) have similar meanings. One can present the joint statistical moments (A.19-A.22) via their averages (A.3; A.4) and correlations (Shiryaev, 1999; Shreve, 2004):

$$CC(t,\beta|1,2) = C(t;1|1)C(t-\beta;1|2) + corr\{C(t|1), C(t-\beta|2)\} \quad (A.23)$$
$$UU(t,\beta|1,2) = U(t;1|1)U(t-\beta;1|2) + corr\{U(t|1), U(t-\beta|2)\} \quad (A.24)$$
$$CU(t,\beta|1,2) = C(t;1|1)U(t-\beta;1|2) + corr\{C(t|1), U(t-\beta|2)\} \quad (A.25)$$
$$UC(t,\beta|1,2) = U(t;1|1)C(t-\beta;1|2) + corr\{U(t|1), C(t-\beta|2)\} \quad (A.26)$$

Now it is possible to calculate $E_m[\delta p(t_i,\beta|1,2)]=corr\{p(t|1),p(t-\beta|2)\}$ (A.17) as an average over the weight function $w(t_i,\beta|1,2)$ (A.14). Taking into account the definition of $\delta p(t_i,\beta|1,2)$ (A.16), the form of the weight function $w(t_i,\beta|1,2)$ (A.14), and equations (A.7; A.8), obtain:

$$corr\{p(t|1), p(t-\beta|2)\} = pp(t,\beta|1,2) - a(t-\beta|2)p(t,\beta|1) - a(t|1)p(t,\beta|2) +$$
$$a(t|1)a(t-\beta|2) \quad (A.27)$$

The function $pp(t,\beta|1,2)$ in (A.27) is the average product of prices $p(t_i|1)p(t_i-\beta|2)$ of assets 1 and 2 over the weight function $w(t_i,\beta|1,2)$ (A.14):

$$pp(t,\beta|1,2) = \sum_{i=1}^{N} p(t_i|1)p(t_i-\beta|2)w(t_i,\beta;1,2) = \frac{1}{U_{\Sigma}(t,\beta|1,2)}\sum_{i=1}^{N} p(t_i|1)p(t_i-\beta|2)U(t_i|1)U(t_i-\beta|2) = \frac{1}{U_{\Sigma}(t,\beta|1,2)}\sum_{i=1}^{N} C(t_i|1)C(t_i-\beta|2)$$

$$pp(t,\beta|1,2) = \frac{1}{U_{\Sigma}(t,\beta|1,2)}\sum_{i=1}^{N} C(t_i|1)C(t_i-\beta|2) = \frac{CC_{\Sigma}(t,\beta|1,2)}{UU_{\Sigma}(t,\beta|1,2)} = \frac{CC(t,\beta|1,2)}{UU(t,\beta|1,2)} \quad (A.28)$$

The function $p(t,\beta|1)$ is the average price $p(t_i|1)$ over the weight function $w(t_i,\beta|1,2)$ (A.14):

$$p(t,\beta|1) = \sum_{i=1}^{N} p(t_i|1)w(t_i,\beta|1,2) = \frac{1}{UU_{\Sigma}(t,\beta|1,2)}\sum_{i=1}^{N} p(t_i|1)U(t_i|1)U(t_i-\beta|2) = \frac{1}{UU_{\Sigma}(t,\beta|1,2)}\sum_{i=1}^{N} C(t_i|1)U(t_i-\beta|2)$$

$$p(t,\beta|1) = \frac{1}{UU_{\Sigma}(t,\beta|1,2)}\sum_{i=1}^{N} C(t_i|1)U(t_i-\beta|2) = \frac{CU_{\Sigma}(t,\beta|1,2)}{UU_{\Sigma}(t,\beta|1,2)} = \frac{CU(t,\beta|1,2)}{UU(t,\beta|1,2)} \quad (A.29)$$

The function $p(t,\beta|2)$ is the average price $p(t_i-\beta|2)$ over the weight function $w(t_i,\beta|1,2)$ (A.14):

$$p(t,\beta|2) = \sum_{i=1}^{N} p(t_i - \beta|2)w(t_i,\beta|1,2) = \frac{1}{U_\Sigma(t,\beta|1,2)} \sum_{i=1}^{N} U(t_i|1)p(t_i - \beta|2)U(t_i - \beta|2) = \frac{1}{U_\Sigma(t,\beta|1,2)} \sum_{i=1}^{N} U(t_i|1)C(t_i - \beta|2)$$

$$p(t,\beta|2) = \frac{1}{U_\Sigma(t,\beta|1,2)} \sum_{i=1}^{N} U(t_i|1)C(t_i - \beta|2) = \frac{UC_\Sigma(t,\beta|1,2)}{UU_\Sigma(t,\beta|1,2)} = \frac{UC(t,\beta|1,2)}{UU(t,\beta|1,2)} \quad (A.30)$$

Using (A.28-A.30), we present the correlation *corr{p(t|1),p(t-β|2)}* (A.27) as:

$$corr\{p(t|1), p(t-\beta|2)\} = \frac{1}{UU(t,\beta|1,2)} [\, CC(t,\beta|1,2) - a(t|1)\, UC(t,\beta|1,2) - a(t-\beta|2)CU(t,\beta|1,2)] + a(t|1)a(t-\beta|2) \quad (A.31)$$

Using (A9; A.10) and (A.23-A.26), one can present the right-hand side (A.31) as:

$$CC(t,\beta|1,2) = C(t;1|1)C(t-\beta;1|2) + corr\{C(t|1), C(t-\beta|2)\}$$

$$a(t|1)\, UC(t,\beta|1,2) = \frac{C(t;1|1)}{U(t;1|1)} [U(t;1|1)C(t-\beta;1|2) + corr\{U(t|1), C(t-\beta|2)\}] = C(t;1|1)C(t-\beta;1|2) + a(t|1)corr\{U(t|1), C(t-\beta|2)\}$$

$$a(t-\beta|2)CU(t,\beta|1,2) = \frac{C(t-\beta;1|2)}{U(t-\beta;1|2)} [C(t;1|1)U(t-\beta;1|2) + corr\{C(t|1), U(t-\beta|2)\}] = C(t;1|1)C(t-\beta;1|2) + a(t-\beta;1|2)corr\{C(t|1), U(t-\beta|2)\}$$

The use of the above relations allow present the correlation *corr{p(t|1)p(t-β|2)}* (A.31) as:

$$corr\{p(t|1), p(t-\beta|2)\} = \frac{1}{UU(t,\beta|1,2)} [corr\{C(t|1), C(t-\beta|2)\} - a(t|1)corr\{U(t|1), C(t-\beta|2)\} - a(t-\beta|2)corr\{C(t|1), U(t-\beta|2)\} + a(t|1)a(t-\beta|2)corr\{U(t|1), U(t-\beta|2)\}] \quad (A.32)$$

The function *UU(t,β|1,2)* (A.24) in the denominator of (A.32) has the form:

$$UU(t,\beta|1,2) = U(t;1|1)U(t-\beta;1|2) + corr\{U(t|1), U(t-\beta|2)\}$$

Finally, the market-based correlation *corr{p(t|1),p(t-β|2)}* (A.32) between prices of two assets depends upon correlations of their trade values and volumes (A.23-A.26), on average trade volumes *U(t;1|1)* (A.3) and *U(t-β;1|2)* (A.4) of two assets, and on VWAP or market-based average prices *a(t|1)* (A.9) and *a(t-β|2)* (A.10).

If assets 1 and 2 are the same, then (A.32) describes correlation of prices of asset 1 (A.33):

$$corr\{p(t|1), p(t-\beta|1)\} = \frac{1}{UU(t,\beta|1)} [corr\{C(t|1), C(t-\beta|1)\} - a(t|1)corr\{U(t|1), C(t-\beta|1)\} - a(t-\beta|1)corr\{C(t|1), U(t-\beta|1)\} + a(t|1)a(t-\beta|1)corr\{U(t|1), U(t-\beta|1)\}] \quad (A.33)$$

Finally, if the time shift between prices *β=0*, then (A.33) equals to market-based volatility $\sigma_p^2(t)$ (A.34; A.35) of price at time *t*:

$$E_m\left[\left(p(t_i|1) - a(t|1)\right)^2\right] = \sigma_p^2(t) \quad (A.34)$$

$$\sigma_p^2(t) = \frac{1}{U(t;2|1)} [\, \Omega_C^2(t) - 2a(t|1)corr\{C(t|1), U(t|1)\} + a^2(t|1)\Omega_U^2(t)] \quad (A.35)$$

Functions $\Omega_C^2(t)$ and $\Omega_U^2(t)$ (A.36) determine volatilities of trade values *C(t_i|1)* and trade volumes *U(t_i|1)* of asset 1 during the averaging interval *Δ* (A.1):

$$\Omega_C^2(t) = C(t;2|1) - C^2(t|1) \quad ; \quad \Omega_U^2(t) = U(t;2|1) - U^2(t|1) \tag{A.36}$$

$$C(t;2|1) = \frac{1}{N}\sum_{i=1}^{N} C^2(t_i|1) \quad ; \quad U(t;2|1) = \frac{1}{N}\sum_{i=1}^{N} U^2(t_i|1) \tag{A.37}$$

Functions *C(t;2|1)* and *U(t;2|1)* in (A.36; A.37) have the meaning of the *2nd* statistical moments or the average squares of trade values and volumes. The relation (A.35) coincides with the expression of price volatility that was derived by Olkhov (2022a).

The expression of correlation *corr{p(t|1),p(t-β|2)}* (A.32) determines the joint market-based statistical moment of the product of prices of two assets *a(t,β|1,2)*:

$$E_m[\,p(t_i|1)p(t_i - \beta|2)] = a(t,\beta|1,2) \tag{A.38}$$

$$a(t,\beta|1,2) = a(t|1)a(t-\beta|2) + corr\{p(t|1), p(t-\beta|2)\} \tag{A.39}$$

$$a(t,\beta|1,2) = \frac{1}{UU(t,\beta|1,2)}[CC(t,\beta|1,2) - a(t|1)corr\{U(t|1), C(t-\beta|2)\} - a(t-\beta|2)corr\{C(t|1), U(t-\beta|2)\} + 2a(t|1)a(t-\beta|2)corr\{U(t|1), U(t-\beta|2)\}] \tag{A.40}$$

If *β=0*, assets 1 and 2 are the same, the joint market-based statistical moment of the product of prices of two assets *a(t,β|1,2)* (A.39) takes the form (A.41):

$$E_m[p^2(t_i|1)] = a(t;2|1) = \frac{C(t;2|1) + 2a^2(t|1)\Omega_U^2(t) - 2a(t|1)corr\{C(t|1), U(t|1)\}}{U(t;2|1)} \tag{A.41}$$

The expression (A.41) coincides with the 2nd market-based statistical moment of price that was derived by Olkhov (2022a). If all trade volumes of assets 1 and 2 are constant during the averaging interval *Δ* (A.1) then the market-based expression of correlation (A.32) takes the form of frequency-based correlation (1.3). Indeed, if *U(t_i|1)=const*, and *U(t_i|2)=const*, then:

$$corr\{U(t|1), C(t-\beta|2)\} = corr\{C(t|1), U(t-\beta|2)\} = corr\{U(t|1), U(t-\beta|2)\} = 0$$

In that case market-based average prices *a(t|1)* (A.9) and *a(t-β|2)* (A.10) equal to frequency-based average:

$$a(t|1) = \frac{\sum_{i=1}^{N} p(t_i|1)U(t_i|1)}{\sum_{i=1}^{N} U(t_i|1)} = \frac{1}{N}\sum_{i=1}^{N} p(t_i|1) = \pi(t|1)$$

$$a(t-\beta|2) = \frac{\sum_{i=1}^{N} p(t_i-\beta|2)U(t_i-\beta|2)}{\sum_{i=1}^{N} U(t_i-\beta|2)} = \frac{1}{N}\sum_{i=1}^{N} p(t_i-\beta|2) = \pi(t-\beta|2)$$

and correlation (A.32) takes the form:

$$corr\{p(t|1)p(t-\beta|2)\} = \frac{corr\{C(t|1), C(t-\beta|2)\}}{UU(t,\beta|1,2)}$$

$$corr\{p(t|1), p(t-\beta|2)\} = \frac{1}{N}\sum_{i=1}^{N}[p(t_i|1) - \pi(t|1)][p(t_i - \beta|2) - \pi(t-\beta|2)] \tag{A.42}$$

Thus, if all trade volumes of assets 1 and 2 are constant, then the correlation between the prices of two assets (A.32) equals the frequency-based correlation (A.42; 1.3).

**Appendix B**

**Market-Based Correlations of Returns of Two Assets**

Let us consider returns $r(t_i,\alpha|1)$ of asset 1 and returns $r(t_i,\beta|2)$ of asset 2:

$$r(t_i,\alpha|1) = \frac{p(t_i|1)}{p(t_i-\alpha|1)} \quad ; \quad r(t_i,\beta|2) = \frac{p(t_i|2)}{p(t_i-\beta|2)} \tag{B.1}$$

The time intervals $\alpha$ and $\beta$ determine the time shift of price $p(t_i\text{-}\alpha|1)$ of asset 1 and the time shift of price $p(t_i\text{-}\beta|2)$ of asset 2 with respect to time $t_i$ that belongs to the averaging interval $\Delta$ (A.1). We use (B.1) and transfer price equations (A.7; A.8) into return equations (B.2; B.3):

$$C(t_i|1) = p(t_i|1)U(t_i|1) = \frac{p(t_i|1)}{p(t_i-\alpha|1)}p(t_i-\alpha|1)U(t_i|1)$$

$$C(t_i|1) = r(t_i,\alpha|1)C_0(t_i,\alpha|1) \quad ; \quad C_0(t_i,\alpha|1) = p(t_i-\alpha|1)U(t_i|1) \tag{B.2}$$

$$C(t_i|2) = r(t_i,\beta|2)C_o(t_i,\beta|2) \quad ; \quad C_o(t_i,\beta|2) = p(t_i-\beta|2)U(t_i|2) \tag{B.3}$$

Function $C_0(t_i,\alpha|1)$ has meaning of the past value of the current trade volume $U(t_i|1)$ of asset 1 in the past trade at time $t_i\text{-}\alpha$ at price $p(t_i\text{-}\alpha|1)$. Function $C_0(t_i,\beta|2)$ has similar meaning of the past value of current trade volume $U(t_i|2)$ of asset *2* in the past at time $t_i\text{-}\beta$ at price $p(t_i\text{-}\beta|2)$. Equations (B.2; B.3) have the form similar to equations (A.7; A.8). That allows us to use the same method to derive the correlations $corr\{r(t,\alpha|1)r(t,\beta|2)\}$ between returns (B.1) of two assets. We introduce the statistical moments and correlations of the past values $C_0(t_i,\alpha|1)$ of the asset 1 and $C_0(t_i,\beta|2)$ of the asset 2 as substitutions of similar statistical moments of the trade volumes $U(t_i|1)$ and $U(t_i|2)$. We define the average past value $C_0(t,\alpha|1)$ and the sum of past values $C_{0\Sigma}(t,\alpha|1)$ (B.4) of the asset 1 and the same functions (B.5) for the asset 2.

$$C_o(t,\alpha|1) = \frac{1}{N}\sum_{i=1}^{N} C_o(t_i,\alpha|1) \quad ; \quad C_{o\Sigma}(t,\alpha|1) = N \cdot C_o(t,\alpha|1) \tag{B.4}$$

$$C_o(t,\beta|2) = \frac{1}{N}\sum_{i=1}^{N} C_o(t_i,\beta|2) \quad ; \quad C_{o\Sigma}(t,\beta|2) = N \cdot C_o(t,\beta|2) \tag{B.5}$$

Equations (B.2) and (B.3) generate VaWAR (3.9; 3.10) or market-based average returns $h(t,\alpha|1)$ (B.6) and $h(t,\beta|2)$ (B.7) that have form similar to VWAP or market-based average price $a(t|1)$ (A.9) and $a(t,\beta|2)$ (A.10). The derivation is given in Olkhov (2023a):

$$E_m[r(t_i,\alpha|1)] = h(t,\alpha|1) = \frac{\sum_{i=1}^{N} r(t_i,\alpha|1)C_o(t_i,\alpha|1)}{\sum_{i=1}^{N} C_o(t_i,\alpha|1)} = \frac{C_\Sigma(t;1|1)}{C_{o\Sigma}(t,\alpha;1|1)} = \frac{C(t;1|1)}{C_o(t,\alpha;1|1)} \tag{B.6}$$

$$E_m[r(t_i,\beta|2)] = h(t,\beta|2) = \frac{\sum_{i=1}^{N} r(t_i,\beta|2)C_o(t_i,\beta|2)}{\sum_{i=1}^{N} C_o(t_i,\beta|2)} = \frac{C_\Sigma(t;1|2)}{C_{o\Sigma}(t,\beta;1|2)} = \frac{C(t;1|2)}{C_o(t,\beta;1|2)} \tag{B.7}$$

To derive market-based correlations between returns of two assets, we introduce the equation (B.8) as the product of equations (B.2; B.3):

$$C(t_i|1)C(t_i|2) = r(t_i,\alpha|1)r(t_i,\beta|2)C_0(t_i,\alpha|1)C_o(t_i,\beta|2) \tag{B.8}$$

Equation (B.8) is similar to equation (A.12). Like in (A.14; A.15), the equation (B.8) generates the weight function $z(t_i,\alpha,\beta|1,2)$ (B.9):

$$z(t_i, \alpha, \beta|1,2) = \frac{C_0(t_i,\alpha|1)C_o(t_i,\beta|2)}{C_0C_{o\Sigma}(t,\alpha,\beta|1,2)} \quad ; \quad \sum_{i=1}^{N} z(t_i, \alpha, \beta|1,2) = 1 \tag{B.9}$$

$$C_0C_{o\Sigma}(t, \alpha, \beta|1,2) = \sum_{i=1}^{N} C_0(t_i, \alpha|1)C_o(t_i, \beta|2)$$

Like in (A.16), we define the product $\delta r(t_i,\alpha,\beta|1,2)$ (B.10) of variations of returns $r(t_i,\alpha|1)$ and $r(t_i,\beta|2)$ of assets 1 and 2 with respect to their VaWAR (3.9; 3.10) or market-based average values $h(t,\alpha|1)$ (B.6) and $h(t,\beta|2)$ (B.7).

$$\delta r(t_i, \alpha, \beta|1,2) = [r(t_i, \alpha|1) - h(t, \alpha|1)][r(t_i, \beta|2) - h(t, \beta|2)] \tag{B.10}$$

We define market-based $corr\{r(t,\alpha|1)r(t,\beta|2)\}$ of returns of assets 1 and 2 during the averaging interval $\Delta$ (A.1) as the product $\delta r(t_i,\alpha,\beta|1,2)$ (B.10) that is averaged over the weight function $z(t_i,\alpha,\beta|1,2)$ (B.9) and define:

$$corr\{r(t, \alpha|1)r(t, \beta|2)\} = \sum_{i=1}^{N} \delta r(t_i, \alpha, \beta|1,2) \, z(t_i, \alpha, \beta|1,2) \tag{B.11}$$

To calculate the expression (B.11), like in (A.19-A.22), we introduce (B.12-B.15):

$$C_oC_o(t, \alpha, \beta|1,2) = \frac{1}{N}\sum_{i=1}^{N} C_o(t_i, \alpha|1)C_o(t_i, \beta|2) \tag{B.12}$$

$$C_oC_{o\Sigma}(t, \alpha, \beta|1,2) = N \cdot C_oC_o(t, \alpha, \beta|1,2) \tag{B.13}$$

$$CC_o(t, \beta|1,2) = \frac{1}{N}\sum_{i=1}^{N} C(t_i|1)C_o(t_i, \beta|2) \quad ; \quad CC_{o\Sigma}(t, \alpha|1,2) = N \cdot CC_o(t, \alpha|1,2) \tag{B.14}$$

$$C_oC(t, \alpha|1,2) = \frac{1}{N}\sum_{i=1}^{N} C_o(t_i, \alpha|1)C(t_i|2) \quad ; \quad C_oC_{\Sigma}(t, \alpha|1,2) = N \cdot C_oC(t, \alpha|1,2) \tag{B.15}$$

The correlations (B.16-B.18) between the current and past values of market trades are determined similar to (A.23-A.26):

$$C_oC_o(t, \alpha, \beta|1,2) = C_o(t, \alpha|1)C_o(t, \beta|2) + corr\{C_o(t, \alpha|1)C_o(t, \beta|2)\} \tag{B.16}$$

$$CC_o(t, \beta|1,2) = C(t|1)C_o(t, \beta|2) + corr\{C(t|1)C_o(t, \beta|2)\} \tag{B.17}$$

$$C_oC(t, \alpha|1,2) = C_o(t, \alpha|1)C(t|2) + corr\{C_o(t, \alpha|1)C(t|2)\} \tag{B.18}$$

Using (B.9-B.10) we present $corr\{r(t,\alpha|1)r(t,\beta|2)\}$ (B.11) of returns of assets 1 and 2 as:

$$corr\{r(t, \alpha|1)r(t, \beta|2)\} = r(t, \alpha, \beta|1,2) - h(t, \beta|2)r(t, \alpha, \beta|1) - h(t, \alpha|1)r(t, \alpha, \beta|2) +$$
$$h(t, \alpha|1)h(t, \beta|2) \tag{B.19}$$

Functions in the right side of (B.19) have the form:

$$r(t, \alpha, \beta|1,2) = \sum_{i=1}^{N} r(t_i, \alpha|1)r(t_i, \beta|2)z(t_i, \alpha, \beta|1,2) =$$
$$\frac{1}{C_oC_o(t,\alpha,\beta|1,2)}\sum_{i=1}^{N} C(t_i|1)C(t_i|2) = \frac{CC(t|1,2)}{C_oC_o(t,\alpha,\beta|1,2)} \tag{B.20}$$

$$r(t, \alpha, \beta|1) = \sum_{i=1}^{N} r(t_i, \alpha|1)z(t_i, \alpha, \beta|1,2) = \frac{CC_o(t,\beta|1,2)}{C_oC_o(t,\alpha,\beta|1,2)} \tag{B.21}$$

$$r(t, \alpha, \beta|2) = \sum_{i=1}^{N} r(t_i, \beta|2)z(t_i, \alpha, \beta|1,2) = \frac{C_oC(t,\alpha|1,2)}{C_oC_o(t,\alpha,\beta|1,2)} \tag{B.22}$$

Using (B.20-B.22), one can present *corr{r(t,α|1)r(t,β|2)}* (B.19) as follows:

$$corr\{r(t,\alpha|1)r(t,\beta|2)\} = \frac{1}{C_o C_o(t,\alpha,\beta|1,2)}[CC(t,\alpha,\beta|1,2) - h(t,\beta|2)\,CC_o(t,\beta|1,2) -$$
$$h(t,\alpha|1)C_o C(t,\alpha|1,2)] + h(t,\alpha|1)h(t,\beta|2) \quad (B.23)$$

The use of (B.16-B.18) gives the form of *corr{r(t,α|1)r(t,β|2)}* (B.23) that is alike to the form of *corr{p(t|1)p(t-β|2)}* (A.32) of prices of two assets:

$$corr\{r(t,\alpha|1)r(t,\beta|2)\} = \frac{1}{C_o C_o(t,\alpha,\beta|1,2)}\,[corr\{C(t|1)C(t|2)\} -$$
$$h(t,\beta|2)corr\{C(t|1)C_o(t,\beta|2)\} - h(t,\alpha|1)corr\{C_o(t,\alpha|1)C(t|2)\} +$$
$$h(t,\alpha|1)h(t,\beta|2)\,corr\{C_o(t,\alpha|1)C_o(t,\beta|2)\}\,]$$

If assets 1 and 2 are the same, then the relations (B.24) describe autocorrelation *corr{r(t,α)r(t,β)}* (B.25) between returns with time shifts α and β.

$$corr\{r(t,\alpha|1)r(t,\beta|1)\} = \frac{1}{C_o C_o(t,\alpha,\beta|1)}\,[\,\Omega_C^2(t) - h(t,\beta|1)corr\{C(t|1)C_o(t,\beta|1)\} -$$
$$h(t,\alpha|1)corr\{C_o(t,\alpha|1)C(t|1)\} + h(t,\alpha|1)h(t,\beta|1)corr\{C_o(t,\alpha|1)C_o(t,\beta|1)\}] \quad (B.25)$$

If *α=β*, (B.25) equals to volatility of return $\sigma_r^2(t,\alpha)$ (B.26) with time shift *α*:

$$\sigma_r^2(t,\alpha) = \frac{\Omega_C^2(t)+h^2(t,\alpha|1)\Phi^2(t,\alpha)-2h(t,\alpha|1)corr\{C(t|1)C_o(t,\alpha|1)\}}{C_o(t,\alpha;2|1)} \quad (B.26)$$

In (B.25; B.26) function $\Omega_C^2(t)$ denotes volatility (A.38) of trade values *C(t_i|1)* and $\Phi^2(t,\alpha)$ denotes volatility (B.27) of past trade values:

$$\Phi^2(t,\alpha) = C_0(t,\alpha;2|1) - C_0^2(t,\alpha|1) \quad (B.27)$$
$$C_0(t,\alpha;2|1) = \frac{1}{N}\sum_{i=1}^{N} C_0^2(t_i,\alpha|1)$$

The function *C_0(t,α;2|1)* in (B.26; B.27) has the meaning of the 2nd statistical moment or the average squares of past values *C_0(t_i,α|1)* of asset 1. Like in (A.37; A.38), the joint statistical moment *h(t,α,β|1,2)* (B.28; B.29) of returns of two assets 1 and 2 takes the form (B.30):

$$h(t,\alpha,\beta|1,2) = E_m[\,r(t,\alpha|1)r(t,\beta|2)] \quad (B.28)$$
$$h(t,\alpha,\beta|1,2) = h(t,\alpha|1)h(t,\beta|2) + corr\{r(t,\alpha|1)r(t,\beta|2)\} \quad (B.29)$$
$$h(t,\alpha,\beta|1,2) = \frac{1}{C_o C_o(t,\alpha,\beta|1,2)}\,[C_o C_o(t,\alpha,\beta|1,2) - h(t,\beta|2)corr\{C(t|1)C_o(t,\beta|2)\} -$$
$$h(t,\alpha|1)corr\{C_o(t,\alpha|1)C(t|2)\} + 2h(t,\alpha|1)h(t,\beta|2)\,corr\{C_o(t,\alpha|1)C_o(t,\beta|2)\}\,] \quad (B.30)$$

If *α=β* and asset 1 is the same as asset 2 then *h(t,α,β|1,2)* (B.30) takes the form:

$$h(t,\alpha,\alpha|1,1) = E_m[\,r^2(t,\alpha|1)] = \frac{C(t;2|1)+2h^2(t,\alpha|1)\Phi^2(t,\alpha)-2h(t,\alpha|1)corr\{C(t|1)C_o(t,\alpha|1)\}}{C_o(t,\alpha;2|1)}$$

If all past values *C_0(t_i,α|1)* of asset 1 and past values *C_0(t_i, β|2)* of asset 2 are constant during *Δ* (1.1), then the correlation (B.24) between returns of assets equals to the frequency-based correlation (2.7). To show that, let us take:

$$C_o(t_i, \alpha|1) = C_o(1) - const \; ; \quad C_o(t_i, \beta|2) = C_o(2) - const$$

$$corr\{C(t|1)C_o(t,\beta|2)\} = corr\{C_o(t,\alpha|1)C(t|2)\} = corr\{C_o(t,\alpha|1)C_o(t,\beta|2)\} = 0$$

Market-based average returns (B.6) and (B.7) take the form of frequency-based:

$$h(t, \alpha|1) = \frac{\sum_{i=1}^{N} r(t_i, \alpha|1) C_o(t_i, \alpha|1)}{\sum_{i=1}^{N} C_o(t_i, \alpha|1)} = \frac{1}{N} \sum_{i=1}^{N} r(t_i, \alpha|1) = \varrho(t, \alpha|1)$$

$$h(t, \beta|2) = \frac{\sum_{i=1}^{N} r(t_i, \beta|2) C_o(t_i, \beta|2)}{\sum_{i=1}^{N} C_o(t_i, \beta|2)} = \frac{1}{N} \sum_{i=1}^{N} r(t_i, \beta|2) = \varrho(t, \beta|2)$$

$$corr\{r(t, \alpha|1) r(t, \beta|2)\} = \frac{corr\{C(t|1)C(t|2)\}}{C_o C_o(t, \alpha, \beta|1,2)}$$

$$C_o C_o(t, \alpha, \beta|1,2) = C_o(1) C_o(2)$$

$$C(t_i|1) - C(t|1) = [r(t_i, \alpha|1) - \varrho(t, \alpha|1)] C_o(1)$$

$$C(t_i|2) - C(t|2) = [r(t_i, \beta|2) - \varrho(t, \beta|2)] C_o(2)$$

$$corr\{C(t|1)C(t|2)\} = C_o(1) C_o(2) C_o corr\{r(t, \alpha|1) r(t, \beta|2)\}$$

$$corr\{r(t, \alpha|1) r(t, \beta|2)\} = \frac{1}{N} \sum_{i=1}^{N} [r(t_i, \alpha|1) - \varrho(t, \alpha|1)][r(t_i, \beta|2) - \varrho(t, \beta|2)]$$

The past values $C_0(t_i, \alpha|1)$ and $C_0(t_i, \beta|2)$ of assets 1 and 2 play a role similar to trade volumes for the description of market-based averages and correlations of prices of two assets.

# Appendix C

## Market-Based Correlations of Prices and Returns of Two Assets

We use results of App. A and App. B to derive correlation $corr\{p(t|1)r(t,\beta|2)\}$ between prices $p(t_i|1)$ (A.7) of asset 1 and returns $r(t_i,\beta|2)$ (B.1; B.3) of asset 2 during the interval $\Delta$ (A.1). We introduce the equation (C.1) as a product of equations (A.7) and (B.3):

$$C(t_i|1) C(t_i|2) = p(t_i|1) r(t_i, \beta|2) U(t_i|1) C_o(t_i, \beta|2) \tag{C.1}$$

Equation (C.1) generates the weight function $\psi(t_i,\beta|1,2)$ (C.2; C.3) that is similar to weight functions $w(t_i,\beta|1,2)$ (A.14) and $z(t_i,\alpha,\beta|1,2)$ (B.9):

$$\psi(t_i, \beta|1,2) = \frac{U(t_i|1) C_o(t_i, \beta|2)}{UC_{o\Sigma}(t, \beta|1,2)} \; ; \quad \sum_{i=1}^{N} \psi(t_i, \beta|1,2) = 1 \tag{C.2}$$

$$UC_o(t, \beta|1,2) = \frac{1}{N} \sum_{i=1}^{N} U(t_i; 1) C_o(t_i, \beta|2) \; ; \; UC_{o\Sigma}(t, \beta|1,2) = N \cdot UC_o(t, \beta|1,2) \tag{C.3}$$

We define the product $\delta pr(t_i,\beta|1,2)$ (C.4) of variations of prices $p(t_i|1)$ and returns $r(t_i,\beta|2)$ with respect to their market-based averages $a(t|1)$ (A.9) and $h(t,\beta|2)$ (B.7):

$$\delta pr(t_i, \beta|1,2) = [p(t_i|1) - a(t|1)][r(t_i, \beta|2) - h(t, \beta|2)] \tag{C.4}$$

We define market-based correlation $corr\{p(t|1)r(t,\beta|2)\}$ (C.5) between prices $p(t_i|1)$ (A.7) and returns $r(t_i,\beta|2)$ (B.1; B.3) as averaging of (C.4) b the weight function $\psi(t_i,\beta|1,2)$ (C.2; C.3):

$$corr\{p(t|1) r(t, \beta|2)\} = \sum_{i=1}^{N} \delta pr(t_i, \beta|1,2) \, \psi(t_i, \beta|1,2) \tag{C.5}$$

To calculate $corr\{p(t|1)r(t,\beta|2)\}$ (C.5) we introduce functions (C.6-C.8):

$$p(t,\beta|1,2) = \frac{1}{UC_{o\Sigma}(t,\beta|1,2)}\sum_{i=1}^{N}p(t_i|1)U(t_i|1)C_o(t_i,\beta|2) = \frac{CC_o(t,\beta|1,2)}{UC_o(t,\beta|1,2)} \quad (C.6)$$

$$r(t,\beta|1,2) = \frac{1}{UC_{o\Sigma}(t,\beta|1,2)}\sum_{i=1}^{N}r(t_i,\beta|2)U(t_i|1)C_o(t_i,\beta|2) = \frac{UC(t|1,2)}{UC_o(t,\beta|1,2)}$$

(C.7)

$$pr(t,\beta|1,2) = \frac{1}{UC_{o\Sigma}(t,\beta|1,2)}\sum_{i=1}^{N}p(t_i|1)r(t_i,\beta|2)U(t_i|1)C_o(t_i,\beta|2) = \frac{CC(t|1,2)}{UC_o(t,\beta|1,2)}$$

(C.8)

The relations (C.6-C.8) allow present $corr\{p(t|1)r(t,\beta|2)\}$ (C.5) as:

$$corr\{p(t|1)r(t,\beta|2)\} = pr(t,\beta|1,2) - a(t|1)r(t,\beta|1,2) - h(t,\beta|2)p(t,\beta|1,2) + a(t|1)h(t,\beta|2)$$

The use of (A.19; A.22; B.17) give:

$$corr\{p(t|1)r(t,\beta|2)\} = \frac{1}{UC_o(t,\beta|1,2)}[corr\{C(t|1)C(t|2)\} -$$

$$h(t,\beta|2)corr\{C(t|1)\,C_o(t,\beta|2)\} - a(t|1)corr\{U(t|1)C(t|2)\} +$$

$$a(t|1)h(t,\beta|2)corr\{U(t|1)C_o(t,\beta|2)\}] \qquad (C.9)$$

If assets 1 and 2 are the same, then $corr\{p(t|1)r(t,\beta|1)\}$ of prices and returns takes the form:

$$corr\{p(t|1)r(t,\beta|1)\} = \frac{1}{UC_o(t,\beta|1)}[\Omega_C^2(t) - h(t,\beta|1)corr\{C(t|1)\,C_o(t,\beta|1)\} -$$

$$-a(t|1)corr\{U(t|1)C(t|1)\} + a(t|1)h(t,\beta|)corr\{U(t|1)C_o(t,\beta|1)\}] \qquad (C.10)$$

If trade volumes $U(t_i|1)$ of asset 1 and past values $C_0(t_i,\beta|2)$ of the asset 2 are constant, then

$$corr\{C(t|1)\,C_o(t,\beta|2)\} = corr\{U(t|1)C(t|2)\} = corr\{U(t|1)C_o(t,\beta|2)\} = 0$$

$$corr\{p(t|1)r(t,\beta|2)\} = \frac{corr\{C(t|1)C(t|2)\}}{UC_o(t,\beta|1,2)}$$

$$corr\{p(t|1)r(t,\beta|2)\} = \frac{1}{N}\sum_{i=1}^{N}[p(t_i|1) - \pi(t|1)][r(t_i,\beta|2) - h(t,\beta|2)]$$

and (C.9) takes the form of frequency-based correlation (4.1-4.3).